\documentclass[11pt,twoside]{article}

\usepackage{asp2006}
\usepackage{epsf}
\usepackage{lscape}

\begin{document}
\title{The Masses of Black Holes in Active Galactic Nuclei}   
\author{Bradley M.\ Peterson}   
\affil{Department of Astronomy, The Ohio State University, 140 West
18th Ave., Columbus, OH 43210, USA}    

\begin{abstract} 
Reverberation mapping methods have been used to measure masses in
about three dozen AGNs. The consistency of the virial masses computed
from line widths and time delays, the relationship between black hole
mass and host-galaxy stellar bulge velocity dispersion, and the
consistency with black hole masses estimated from stellar dynamics in
the two cases in which such determinations are possible all 
indicate that reverberation mass measurements are robust and 
are accurate to typically a factor of a few.
The reverberation-mapped
AGNs are of particular importance because they anchor the 
scaling relationships that allow black hole mass estimation
based on single spectra. We discuss potential sources of systematic error,
particularly with regard to how the emission line widths are measured.
\end{abstract}

\section{Introduction}

It is only in this first decade of the century that we have been able to discuss
the masses of AGN black holes with any confidence.  For a somewhat
longer time, however, there have been plausible measurements of
the masses of  black holes in quiescent galaxies, based on studies of stellar
dynamics and gas dynamics in the nuclei of nearby galaxies, including
our own Milky Way.  The
presence of an active nucleus in a galaxy makes it extremely difficult to study
the nuclear stellar dynamics; the stellar absorption features are
simply washed out by the AGN spectrum. But gas motions on otherwise
unresolvable angular scales can be studied by use of reverberation
mapping methods (Kaspi, these proceedings).
In this review, we discuss progress made in
determining the masses of black holes in AGNs by reverberation mapping
of the broad-line region (BLR).

\section{On the Reliability of Reverberation-Based Mass Measurements}

Three lines of evidence suggest that the black hole masses based on
reverberation are reliable at the ``factor of a few'' level:
\begin{enumerate} 
\item In each case where it can be reliably tested,
there is a virial-like relationship between the reverberation-based
emission-line lags $\tau$ and the emission-line width $\Delta V$ of
the form $\Delta V \propto \tau^{-1/2}$,
as expected if the dynamics of the BLR are dominated by the gravity of
the central black hole. Specifically, the mass the the black hole is
given by 
\begin{equation} 
\label{eq:Bhmass} M_{\rm BH} = f
\frac{\Delta V^2 c \tau}{G}, 
\end{equation} 
where $G$ is the
gravitational constant and $f$ is a dimensionless factor of order
unity that depends on the kinematics, geometry, and inclination of the
AGN.  
\item Reverberation-based black hole masses correlate with the
stellar bulge velocity dispersion of the host galaxies $\sigma_*$ in a
fashion similar to that seen in quiescent galaxies. In other words,
AGNs also obey the well-known $M_{\rm BH}$--$\sigma*$ relationship.
\item In the rare cases where the black hole mass can be measured by
other methods, there is general agreement with the reverberation-based
masses.  
\end{enumerate} 


\subsection{The Virial Relationship} As noted above, the size of the
emitting region for any particular emission line is well correlated
with the emission-line width in the sense that $\Delta V \propto
R^{-1/2}$, as expected if gravity is the principal dynamical force in
the BLR. This has been shown to hold in every case where it can be
reasonably tested (meaning that some reasonable range of $R$ must be
sampled through measurement of different emission lines or the same
emission line in very different flux states). Indeed, the better the
data (not only in terms of quality, but in terms of the range of
$\Delta V$ and $\tau$ sampled), the better the agreement with the
virial prediction. The best agreement is seen in the well-studied case
of NGC 5548 (Peterson \& Wandel 1999; Peterson et al.\ 2004; Bentz et
al.\ 2007), but a similar relationship is seen in NGC 7469 and 3C
390.3 (Peterson \& Wandel 2000), NGC 3783 (Onken \& Peterson 2002), Mrk
110 (Kollatschny 2003a), and NGC 4151 (Metzroth, Onken \& Peterson
2006; Bentz et al.\ 2006b).

This relationship was anticipated in work by Krolik
et al.\ (1991), but agreement with the virial prediction was not strong
enough to convince these authors that indeed the BLR was
virialized. The analysis of Peterson \& Wandel (1999)  differed from that
of Krolik et al.\  by (1) exclusion of blended or contaminated
lines, (2) exclusion of spurious or dubious line lags, (3) measurement
of the line width in the {\em variable} part of the spectrum (as
discussed below) and (4) most importantly it turns out, elimination of
discretization effects arising from use of the
Discrete Correlation Function method (Edelson \& Krolik 1988), which
returned lags only in multiples of 4 days and with dubious
uncertainties. In the interim, extensive testing (e.g., White \&
Peterson 1994; Peterson et al.\ 1998) demonstrated that greater precision
and accuracy could be achieved using the interpolation method as long
as the data are reasonably well sampled.

Two ways of characterizing the emission-line widths have been employed
to date, and we emphasize that these are {\em not}
interchangeable. The simplest and most widely used line-width measure
is the full-width-at-half maximum (FWHM). The other, originally
suggested in this context
by Fromerth \& Melia (2000) and later championed by Peterson
et al.\ (2004), is the line dispersion $\sigma_{\rm line}$, which is
based on the second moment of the emission-line profile\footnote{It
should be recognized that $\sigma_{\rm line}$ is well-defined
regardless of the emission-line profile. It specifically is {\em not},
as sometimes mistakenly asserted, the width of the best-fit
Gaussian to the line profile. It can be measured in mean, rms,
or single-epoch spectra.}. On one hand, FWHM is trivial to
measure and is relatively insensitive to extended wings and blending
with other lines. On the other hand, $\sigma_{\rm line}$ is
well-defined for even noisy profile data (often the case, as noted
below), is less sensitive to the presence of a narrow-line component,
and is more precise for low-contrast lines.

Peterson et al.\ (2004) argue that the virial relationship between
line width and lag is best reproduced by using $\sigma_{\rm line}$
(as opposed to FWHM) and the centroid of the continuum/emission-line
cross-correlation function (CCF) $\tau_{\rm cent}$ (as opposed to the
peak of the CCF $\tau_{\rm peak}$). They also point out that the
virial relationship is more clearly seen by measuring the line width
in the {\em variable} part of the spectrum: the spectra used to
measure the continuum and emission-line light curves can be combined
to form mean and root-mean-square (rms) spectra; the rms spectrum
isolates the part of the emission line that is actually varying, i.e.,
the gas to which the measured lag applies. 

\section{\boldmath The AGN $M_{\rm BH}$--$\sigma_*$ Relationship}

There is  a tight correlation
between the mass of the central black hole $M_{\rm BH}$ and the
stellar bulge velocity dispersion $\sigma_*$ (Ferrarese \& Merritt
2000; Gebhardt et al.\ 2000a), first discovered in quiescent galaxies. It has
been subsequently shown that reverberation-based black hole masses for AGNs
yield a similar relationship  (Gebhardt et al.\ 2000b; 
Ferrarese et al.\ 2001; Onken et al.\ 2004;
Nelson et al.\ 2004). Onken et al.\ (2004)
determine a mean value for the scaling factor $f$ in eq.\
(\ref{eq:Bhmass}) by matching the zero-point of the AGN $M_{\rm
BH}$--$\sigma_*$ relationship to that of quiescent galaxies. With
values of $\sigma_*$ available for approximately half of the
reverberation-mapped sample, Onken et al.\ derived a mean value of the
scaling factor $\langle f \rangle = 5.5 \pm 1.8$. The consistency
between the AGN $M_{\rm BH}$--$\sigma_*$ relationship and that for
quiescent galaxies provides an additional argument that the
reverberation-based masses have some meaning. Moreover, the scatter
around the best-fit $M_{\rm BH}$--$\sigma_*$ relationship suggests
that, on average, the reverberation masses are accurate to better than
0.5 dex.

\subsection{Comparison with Stellar Dynamical Masses}

The black hole ``radius of influence,'' $r_* =
GM_{\rm BH}/\sigma_*^2$, defines the size of the
nucleus in which the gravity of the black hole dominates the dynamics
of the stars in the galactic bulge. Only two reverberation-mapped
AGNs, NGC 3227 and NGC 4151, are close enough that their black hole
radius of influence is in principle resolvable with {\em Hubble Space
Telescope}, a 2.4-m telescope working at the diffraction limit.

In the case of NGC 3227, Davies et al.\ (2006) used $K$-band integral
field data obtained with SINFONI on the ESO VLT to model the stellar
dynamics of the inner bulge of this galaxy. They find that the central
black hole mass is in the range $(7$--$20) \times 10^6\,M_\odot$, which
compares favorably with the reverberation measurement of $(42\pm21)
\times 10^6\,M_\odot$ (Peterson et al.\ 2004), especially given the
systematic uncertainties in both mass determinations.  Onken et al.\
(2007) attempted to observe the Ca\,{\sc ii} triplet in the near IR
spectrum of NGC 4151 using STIS on {\em HST}. Despite elaborate
attempts to reduce the AGN contribution to the spectrum, 
the Ca\,{\sc ii} triplet lines remained undetectable in the {\em HST} data. 
Fortunately, however, it was possible to use two long-slit spectra of NGC 4151
obtained on the NOAO Mayall 4-m telescope on Kitt Peak and the MMT on
Mt.\ Hopkins, and these proved to be good enough to model the
dynamics. But because NGC 4151 is a non-axisymmetric system (it has a
clear central bar) and only two slit positions were obtained, the mass
determination is quite model-dependent. If it is assumed that the
inclination of the bulge is the same as the disk of the galaxy, then
the central black hole mass is in the range $(40$--$50) \times
10^6\,M_\odot$. If, on the other hand, it is assumed that we observe
the bulge edge-on, then we obtain only an upper limit to the black
hole mass, $M_{\rm BH} < 42 \times 10^6\,M_\odot$. Either result
is consistent with the most recent reverberation result, $M_{\rm BH} =
(46 \pm 5) \times 10^6\,M_\odot$ (Bentz et al.\ 2006b).

In both cases, the agreement between the reverberation and stellar
dynamical results is gratifying. However, neither stellar dynamical
mass measurement is particularly robust, and certainly additional
observations with an integral field unit and adaptive optics will be
required to make a truly convincing case.

\begin{figure}
\plotfiddle{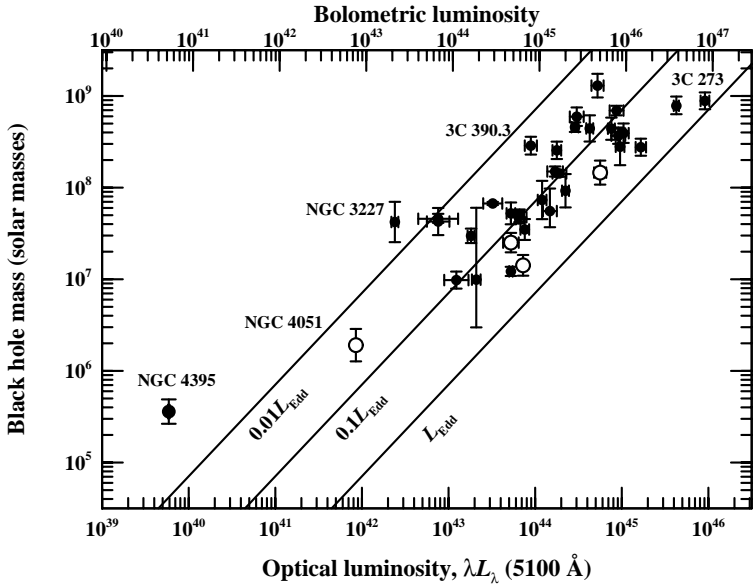}{3.0in}{0}{40}{40}{-148}{0}
\caption{The black hole mass--luminosity relationship for the
re\-ver\-ber\-a\-tion-mapped AGNs. The open circles identify NLS1s,
which are thought to be high Eddington ratio
objects. The bolometric scale on top assumes that
$L_{\rm bol} = 9 \lambda L_{\rm \lambda}\mbox{(5100\,\AA)}$.
This is an updated version of Fig.\ 8 of Peterson et al.\ (2005),
with revised masses for NGC 4151 (Bentz et al.\ 2006b) and NGC 4593
(Denney et al.\ 2006 and these proceedings).}
\end{figure}

\section{The Mass--Luminosity Relationship} 
Based on
reverberation-mapping results for three dozen AGNs, we can begin to
construct a mass--luminosity relationship, as shown in Fig.\ 1.
There are several notable features.
First, reassuringly, all of the AGNs are accreting at
sub-Eddington rates, although in each case we have made the unsupported
general assumption that the bolometric luminosity
is simply a multiple of the optical luminosity, specifically $L_{\rm
bol} = 9 \lambda L_{\lambda}{\mbox{\rm (5100\,\AA)}}$. Second, most of the
AGNs that can be classified as narrow-line Seyfert 1 (NLS1) galaxies
(open circles in the Figure) are found to be high accretion-rate
objects. Moreover, there are several objects with similar ``I Zw 1''
type spectra that do not formally meet the NLS1 criterion of ${\rm
FWHM} < 2000\,{\rm km\ s}^{-1}$, but are nevertheless quite similar
objects\footnote{We submit the following oversimplified argument:
as shown in the next section, the size of the BLR scales like $r
\propto L^{1/2}$. From the virial equation (eq.\ \ref{eq:Bhmass}), the
line width scales like $\Delta V \propto (M/r)^{1/2}$. We define the
Eddington ratio $\dot{m}$ as the accretion rate relative to the
Eddington rate, i.e., $\dot{m} = \dot{M}/\dot{M}_{\rm Edd} \propto
\dot{M}/M$. Combining these, we can write 
$$\Delta V
\propto \left( \frac{M}{L^{1/2}} \right)^{1/2} \propto \left(
\frac{M}{\dot{M}^{1/2}} \right)^{1/2} \propto \left( \frac{M}{\dot{m}}
\right)^{1/4}.  $$
The suggestion is that NLS1 properties
are attributable to the Eddington ratio, i.e., objects with similar
$\dot{m}$ should have similar spectra. However, we see that the line
widths of more massive objects increase with $M$ so that high-mass
AGNs with $\dot{m}$ similar to those of low-mass AGNs do not meet the
NLS1 line-width criterion of ${\rm FWHM} < 2000\,{\rm km\ s}^{-1}$
even though the spectra are otherwise similar. Some of the
higher-luminosity AGNs (e.g., 3C 273 [PG 1226+023] and PG 1700+518) would
be classified with the open-circle objects in Fig.\ 1 if the
criterion was based on $\dot{m}$ rather than line width only.}. Third,
there are a few galaxies that seem to be outliers relative to the
rest of the distribution: specifically, some objects appear to have
abnormally low Eddington ratios. Some of these, like NGC 3227 and NGC
4051, are heavily reddened and have significant internal extinction, so their
luminosities (and hence Eddington ratios) are clearly underestimated.

\section{Estimating Black Hole Masses from Individual Spectra}

The 36 reverberation-mapped AGNs shown in Fig.\ 1 are of
particular importance because they anchor the black hole mass scale,
as we will see in this section.

\subsection{The Radius--Luminosity Relationship} 
There is a remarkable
correlation between the BLR radius ($R \propto c\tau_{\rm cent}$), as
measured for some particular emission line, and the luminosity $L$ of
the AGN. The existence of the correlation itself is not so remarkable;
indeed it was anticipated when there were only a handful of admittedly rather 
unreliable measurements of the BLR size (e.g., Koratkar \& Gaskell 1991),
but appeared only clearly when PG quasars were added to the
reverberation-mapped sample of AGNs (Kaspi et al.\ 2000). It is
usually supposed that it was the extension of the luminosity range to
higher values that led to the emergence of the $R$--$L$ relationship,
but this is only partially correct. The high-luminosity PG quasars
were of particular importance in defining the $R$--$L$ relationship {\em
primarily} because they are so luminous that the luminosity
measurement is affected only a small amount by contamination by the
host galaxy. As showed by Bentz et al.\ (2006a, and these proceedings), the
low-luminosity end of the $R$--$L$ relationship is obscured by the
substantial contamination of the luminosity measurement by starlight
from the host galaxy. When the host-galaxy starlight contribution is
correctly accounted for, the beautiful $R$--$L$ relationship shown by
Bentz et al.\ emerges; over the whole luminosity range, a good fit is
obtained with $R \propto L^{0.51 \pm 0.03}$. So what {\em is}
remarkable about this relationship is: 
\begin{enumerate} \item The
slope of the $R$--$L$ relationship is consistent with the most
na\"{\i}ve expectation if  the physical
conditions in the BLR are the same in all AGNs. This is 
a low-order approximation that flies in the face of
observations (e.g., the Baldwin Effect) and theory (e.g., more massive
black holes should have cooler accretion disks, and hence the shape of
the ionizing continuum should be a function of black hole mass).
\item The {\em starlight-corrected} $R$--$L$ relationship now
appears to have so little scatter that outliers are suspicious: either
their lags are dubious or their luminosities are underestimated
because of internal reddening, for example.  \end{enumerate}

\subsection{Using (or Abusing) Scaling Relationships}

The beauty of the radius--luminosity relationship is that it allows us
to estimate the size of the line-emitting region by measurement of the
luminosity alone. By combining this with the line width, we can then
estimate the mass of an AGN from a single spectrum (Wandel,
Peterson, \& Malkan 1999; Vestergaard 2002, 2004;  McLure \& Jarvis
2002; Kollmeier et al.\ 2006; Vestergaard \& Peterson 2006). This has
to be done carefully, however, to avoid introducing systematic
errors. In the case of H$\beta$, the narrow-line component of H$\beta$ needs to be
accounted for. Blending with other lines, especially Fe\,{\sc ii}, has
to be dealt with.  Also, it is important to realize that the scale
factor $f$ in eq.\ (\ref{eq:Bhmass}) has to be appropriate for the
line-width measure and type of spectrum being measured:
\begin{enumerate} 
\item As already noted, FWHM and $\sigma_{\rm line}$
cannot be used interchangeably; to do so introduces a bias in the
mass scale, as will be described below.  
\item The scale factor $f$ that should be used for estimating the black hole
mass from a single spectrum is {\em not} necessarily the factor $5.5$
determined by Onken et al.\ (2004); this value calibrates
the black hole mass scale based on measuring $\sigma_{\rm line}$ in rms
spectra. Typically, the variable part of the H$\beta$ line is about
20\% {\em narrower} than the whole, uncontaminated line, thus
requiring a different scale factor.
\end{enumerate}

The line-width ratio ${\rm FWHM}/\sigma_{\rm line}$ is a
crude parameterization of the line profile. It is trivial to show that
this ratio has a value of $2(\ln 2)^{1/2} \approx 2.35$ for a Gaussian
profile, which provides a good benchmark for comparison with
observed values. Values of ${\rm FWHM}/\sigma_{\rm line} > 2.35$ are
found for profiles that are more rectangular or ``boxy'' than a
Gaussian; values ${\rm FWHM}/\sigma_{\rm line} < 2.35$ describe
profiles that have narrower cores and broader wings than a Gaussian,
i.e., they are ``peakier'' than a Gaussian. The observed values of
${\rm FWHM}/\sigma_{\rm line}$ for the mean spectra of the
reverberation-mapped AGNs range from $\sim0.71$ (PG 1700+518) to
$\sim3.45$ (Akn 120), with an average value of $\sim 2.0$. Lower
values are found for objects that can be classified as NLS1s or
similar I Zw 1-type objects --- in other words, the line width ratio
seems to correlate with other properties, specifically accretion rate,
although rather imperfectly (Collin et al.\ 2006). The important thing
to note, however, is that simply substituting, say, FWHM for
$\sigma_{\rm line}$ will systematically change the black hole mass
scale for NLS1s relative to the mass scale for other
reverberation-mapped AGNs.

This, of course, doesn't tell us which line width measure is the best
one to use, i.e., the one that introduces less bias relative to the
``true'' black hole masses. In an attempt to identify the best
line-width estimator, Collin et al.\ (2006) performed a simple
test. They divide the sample of reverberation-mapped AGNs into two
subsamples based on the line-width ratio, a Population 1 with ${\rm
FWHM}/\sigma_{\rm line} < 2.35$ and a Population 2 with ${\rm
FWHM}/\sigma_{\rm line} > 2.35$; using the Gaussian value as the
separator puts about half of the AGNs for which $\sigma_*$
measurements are available in each subsample, thus enabling a separate
determination of the scaling factor in eq.\
(\ref{eq:Bhmass}). Repeating the analysis of Onken et al.\ (2004) for
the subsamples, using $\sigma_{\rm line}$ as the line-width measure,
Collin et al.\ (2006) find that the scale factors for Populations 1 and
2 are nearly identical.  On the other hand, when FWHM
is used as the line-width measure, the scale factors for the two
populations are quite different. They then
separate the reverberation-mapped sample
into two subsets on the basis of line width alone, and divide the
reverberation-mapped sample into a Population A with $\sigma_{\rm
line} < 2000$\,km s$^{-1}$\ and Population B with $\sigma_{\rm line} >
2000$\,km s$^{-1}$ (cf.  Sulentic et
al.\ 2000).  Using $\sigma_{\rm line}$ for the line-width measure,
Collin et al.\ find consistent scale factors. And again, when
FWHM is used as the line measure, the scale factors are inconsistent
with one another.  It is thus concluded that using $\sigma_{\rm line}$ as
the line-width measure in eq.\ (\ref{eq:Bhmass}) gives a less-biased
mass scale than using FWHM because the scale factors for the former do
not depend as strongly on either line width or profile.

It is worth repeating that, like all work on the scale factor $f$ thus
far, the calibration is statistical in nature and does not necessarily
apply to individual sources. These statistically determined scaling
factors are intended to give correct mean masses for a sample of
objects: the results should be unbiased in the sense that we
overestimate as many masses as we underestimate.

\subsection{Can We Determine the Inclination of the BLR?}

As discussed by Collin et al.\ (2006), there is evidence that both
${\rm FWHM}/\sigma_{\rm line}$ and $\sigma_{\rm line}$ are affected by
both Eddington ratio and inclination of the BLR. Inclination is
potentially the single largest systematic effect in measuring masses:
the mass inferred will be underestimated by a factor of $1/\sin^2 i$
(where $i$ is the inclination), which is especially worrisome since
unification suggests that we should see most AGNs at low inclination,
at $0\deg \leq i \leq i_{\rm max}$, where $i_{\rm
max}$ is the opening angle of the obscuring torus, probably in the
range $45\deg$--$60\deg$.

It has been suggested by several authors (Wu \& Han 2001; Zhang \& Wu
2002; McLure \& Dunlop 2001) that if inclination is the major
systematic factor affecting reverberation masses, it might be possible
to infer the inclination of the BLR by comparing the
reverberation-based mass (henceforth $M_{\rm rev}$) with some other
estimator, such as the mass predicted by the $M_{\rm BH}$--$\sigma_*$
relationship (which we will refer to henceforth as $M_{\sigma_*}$). We
would expect, for example, that at least some NLS1s or other likely
low-inclination sources would have unusually small values of the ratio
$M_{\rm rev}/M_{\sigma_*}$.  As a first step in investigating this,
let us consider a couple of illuminating test cases.

\paragraph{\bf 3C 120:} This is a particularly interesting source in
this context because its inclination is constrained to be close to
face-on ($i < 20\deg$) by the existence of a superluminal jet
(Marscher et al.\ 2002). At an inclination of $20\deg$, the width of a
line arising from a thin disk is so large that the reverberation-based
mass ought to underpredict the true mass by an order of magnitude. This
source, however, does not stand out in any way in the $M_{\rm
BH}$--$\sigma_*$ relationship.

\paragraph{\bf Mrk 110:} This  NLS1 has an independent mass
estimate based on the gravitational redshift of its strong emission
lines (Kollatschny 2003b). The reverberation mass is $M_{\rm rev} = 25\,
(\pm 6) \times 10^6\,M_\odot$, which is slightly larger than that
predicted by the gravitational redshift, $M_{\rm grav} = 14\,(\pm 3)
\times 10^6\,M_\odot$;  of course, given that a mean value of $f$
was used to compute the reverberation-based mass, these values are perfectly
consistent. In contrast, the prediction of the $M_{\rm
BH}$--$\sigma_*$ relationship is much smaller, $M_{\sigma_*} = 4.8
\times 10^6\,M_\odot$. If the narrowness of the emission lines
was due to inclination alone, we would expect $M_{\sigma_*}$ to be larger than
$M_{\rm rev}$ and closer to $M_{\rm grav}$ since 
neither $M_{\rm grav}$ nor
$M_{\sigma_*}$ are expected to be inclination-dependent.

\paragraph{NGC 4151:} As noted above, the reverberation and stellar
dynamical masses of NGC 4151 are consistent at $\sim 40 \times
10^6\,M_\odot$. The predicted mass from the $M_{\rm BH}$--$\sigma_*$
relationship is an {\em order of magnitude smaller} than this,
$M_{\sigma_*} \approx 4 \times 10^6\,M_\odot$.

\paragraph{}
These three examples all run quite counter to the expected trend if
small line widths are attributable primarily to inclination
effects. However, Collin et al.\ (2006) make a statistical argument
that some objects with low values of ${\rm FWHM}/\sigma_{\rm line}$
could be low-inclination sources if there is a vertical component
(i.e., along the disk axis) that also contributes to the line width: a
contribution of 10--30\% of the disk-plane
velocity dispersion in the $z$-direction
would be consistent with the distribution of
observed ${\rm FWHM}/\sigma_{\rm line}$ values.

Since $M_{\sigma_*}/M_{\rm rev}$ doesn't appear to be a very good predictor of
inclination, we need to consider other ways to obtain even approximate
measures of inclination of the BLR. Candidates include radio jets (as
in the case of 3C 120), spectropolarimetry of the
broad emission lines (see Lira, these proceedings),
and velocity-resolved reverberation mapping, since a high-fidelity
velocity--delay map should allow us to infer the inclination of the
BLR directly.

\section{Future Progress}

As noted above, the $M_{\rm BH}$--$\sigma_*$ relationship does not
appear to be a good enough predictor of AGN black hole mass to infer
the inclination of the BLR in a particular AGN
from $M_{\sigma_*}/M_{\rm rev}$.
However, Collin et al.\ (2006) find at least a suggestion that
the line-width ratio ${\rm FWHM}/\sigma_{\rm line}$ might be
statistically a good predictor. Whether or not $M_{\sigma_*}/M_{\rm rev}$ is a
{\em statistically} good predictor as well (which amounts to
asking whether inclination effects are even important) remains to be
seen: at the present time there are simply far too few sources for
which $\sigma_*$ measurements are available. The prospects for getting
additional measurements of $\sigma_*$ are unclear;
all of the ``easy'' sources have been done. The
reverberation-mapped AGNs for which $\sigma_*$ measurements are
unavailable are those at redshifts $z > 0.06$ where the
Ca\,{\sc ii} triplet is  redshifted into the near-IR water vapor bands.
Moreover, the bulge light is
a proportionally smaller contributor to the total light, 
and the bulge is spatially less well-resolved in these more luminous,
more distant AGNs, compounding the difficulty of observing the
weak absorption features. We can, however, use redshifted CO
transitions in the $H$-band to measure bulge velocity dispersions, and
this has been done in a few cases (e.g., Dasyra et al.\ 2006). However, for
sources with both CO and Ca\,{\sc ii} measurements, the agreement
between them is not as good as one would hope, 
perhaps only because of the relatively low signal-to-noise ratios of
the IR spectra at this time. Of course, prospects for improving the quality of the
IR spectra are quite good as integral field units and laser adaptive
optics systems are deployed on telescopes like VLT and Gemini.

Another crucial need is to obtain at least one high-fidelity 
velocity--delay map for at least one emission line in one AGN
to serve as 
proof of concept. Based on years of experience in studing
emission-line variability, Horne et al.\ (2004) carried out
simulations that underscore the need for high signal-to-noise ratio
($S/N$), highly homogeneous, moderate-resolution spectra, high time
resolution, and a sufficiently long duration program. For typical
bright Seyfert galaxies, it seems that it would be possible to obtain
a good velocity-delay map of H$\beta$ or H$\alpha$ based on spectra
with $S/N$ in the range 30--100 and spectral resolution $\leq
600$\,km s$^{-1}$, sampled once per day for several months. While this sounds
like a fairly tall order, comparison with previous
reverberation-mapping campaigns reveals that this is not much beyond
what has been done already: indeed, some of the better previous
programs met nearly all of the criteria outlined above.

\section{Conclusions} Good progress has been made 
in using reverberation mapping to measure BLR
radii and corresponding black hole mases.  Black hole masses have been
measured for some three dozen AGNs; based on their
scatter around the $M_{\rm BH}$--$\sigma_*$ relationship,
reverberation-based masses appear to be accurate to a factor of about
three. However, we find that in individual cases, the
$M_{\rm BH}$--$\sigma_*$ relationship is not a particularly good
predictor of black hole mass, much less inclination. We are continuing
both direct and statistical tests to determine the accuracy of
reverberation-based masses and to identify systematic
biases.

The relatively few AGNs for which we have reverberation-based masses
are a precious commodity because the data are so hard-won and because
these are the direct measurements that  anchor the AGN black-hole
mass scale. The relationship between BLR radius and AGN continuum
luminosity (Bentz et al., these proceedings) is now sufficiently
well-defined, at least for the Balmer lines, that we can use the
predictions with considerable confidence, provided of course that we
use the correct factor $f$ to compute the masses. At the present time, it
appears that the systematic uncertainties in masses based on
scaling relationships are no worse than a factor of $\sim 4$ (Vestergaard
\& Peterson 2006).

We have also pointed out that the full potential of reverberation
mapping has yet to be realized. We still have not obtained a
high-fidelty velocity--delay map of even a single source, but we
also note that significant improvements are
certainly within reach.

\acknowledgements 
I am grateful for support of
reverberation-mapping studies at Ohio State University by the US
National Science Foundation through grant AST-0604066 and NASA
through grant HST-AR-10691 from STScI.
I thank my
collaborators, particularly M.C.\ Bentz, S.\ Collin, K.D.\ Denney, M.\
Dietrich, T.\ Kawaguchi, C.A.\ Onken, R.W.\ Pogge, and M.\ Vestergaard.

\end{document}